\documentclass{jltp}

\usepackage{lfmath} 
\usepackage{citesort}
\usepackage{amsmath}
\usepackage{amsfonts}
\title{Recent Developments in Understanding  Two-dimensional Turbulence and  the Nastrom-Gage Spectrum}

\author{Eleftherios Gkioulekas and Ka-Kit Tung \address{Department of Applied Mathematics, University of Washington}}

\runninghead{Eleftherios Gkioulekas and Ka-Kit Tung}{two-dimensional and quasi-geostrophic turbulence}
       
\begin{document}

\pagestyle{plain}
\maketitle

\begin{abstract}
Two-dimensional turbulence appears to be a more formidable problem than three-dimensional turbulence despite the numerical advantage of working with one less dimension.  In the present paper we review recent numerical investigations of the phenomenology of two-dimensional turbulence as well as recent theoretical breakthroughs by various  leading researchers. We also review efforts to reconcile the observed energy spectrum of the atmosphere (the spectrum) with the predictions of two-dimensional turbulence and quasi-geostrophic turbulence.

\end{abstract}


\section{Introduction}

Turbulence is ubiquitous in the fluid enviroment we live in, and yet a fundamental theoretical understanding from first principles   is not yet available, although considerable progress has been made in the case of isotropic and homogeneous three-dimensional turbulence.  Large-scale flows in thin fluid shells, such as planetary atmospheres and the ocean,  tend to be quasi-two-dimensional. 
Two-dimensional flows differ from three-dimensional turbulence   in that there are usually two closely related conservative quantities exchanged by nonlinear triad interactions. Furthermore, the cascades of two-dimensional turbulence do not exhibit universal behavior with the same degree of consistency that we have come to expect from three-dimensional turbulence. Also  interesting is the inverse energy cascade, unique in ``2d-like'' systems, where an initially noisy velocity field continuously forced by white noise small-scale forcing will nonetheless evolve into coherent vortical structures. The striking resemblence between the pattern formation of two-dimensional turbulence and similar patterns in the atmospheres of gas-giant planets, like Jupiter, tickles the imagination and raises interesting but hard questions. \cite{article:Yano:1994}

When Kraichnan \cite{article:Kraichnan:1967:1}, Leith \cite{article:Leith:1968} and Batchelor \cite{article:Batchelor:1969} first  pioneered the study of two-dimensional turbulence, it was thought that  it would be easier to handle theoretically and simpler to simulate numerically than three-dimensional turbulence.  The  fact that no convincing simulation of the  dual cascades predicted by KLB, with  an upscale energy cascade and a downscale enstrophy cascade, has been  achieved during the ensuing three decades is a hint that the problem of  two-dimensional turbulence is richer than was thought, perhaps even  richer than the three-dimensional isotropic homogeneous turbulence.  In  addition, because   geophysical fluids behave more like  two-dimensional fluids than three-dimensional isotropic homogeneous  fluids, it is not possible to simply ignore the theoretical and numerical  problems of two-dimensional turbulence on the grounds that it is a fictitious fluid. 

In the present paper, we shall review some of the recent breakthroughs in understanding two-dimensional turbulence.  We shall also review the problem of the Nastrom-Gage energy spectrum of the atmosphere, and recent theories that have been proposed to explain it.  Needless to say, this review is biased to reflect the viewpoint and  interests of the authors.  Less biased reviews of two-dimensional turbulence \cite{article:Montgomery:1980,book:Lesieur:1990,article:Tabeling:2002} and quasi-geostrophic turbulence \cite{book:Pedlosky:1979,book:Salmon:1998} are available in the literature. Good reviews on the Nastrom-Gage spectrum can also be found in the papers by Lindborg \cite{article:Lindborg:1999,article:Lindborg:2006}. 

This paper is organized as follows. Sections 2 and 3 of the paper deal with two-dimensional turbulence.  Section 4 discusses the problem of the Nastrom-Gage  energy spectrum.  Finally, section 5 reviews the method of spectral reduction, because we believe that it has the potential to lead to further breakthroughs in this field.

\section{Dynamics of  Two-dimensional Turbulence}

 Let  $u_{\alpha}(\bfr, t)$  be the Eulerian velocity field.  The governing equations  of two-dimensional turbulence are:
\begin{align}
 \pderiv{u_{\ga}}{t} &+ u_{\gb}\partial_{\gb}u_{\ga} = -\partial_{\ga}p + \cD u_\ga + f_{\ga},\\
\partial_{\ga}u_{\ga} &= 0,
\end{align}
where $f_\ga$ is the forcing term, and $\cD$ is the dissipation operator  given by
\begin{equation}
\cD  \equiv (-1)^{\gk+1} \nu_{\gk}\del^{2\gk}   + (-1)^{m+1}\gb \del^{-2m}.
\end{equation}
Here the integers $\gk$ and $m$ describe the order of the dissipation mechanisms, and  the numerical coefficients $\nu_{\gk}$ and $\gb$ are the corresponding viscosities.  $\cD$ is the overall dissipation operator.   The term $f_\ga$ represents stochastic forcing that injects energy into the system at a range of length scales in the neighborhood of the integral length scale $\ell_0$. The term $\gb \del^{-2m} u_{\ga}$ describes a dissipation mechanism that operates on large-scale motions. The operator $\del^{-2m}$ represents applying the inverse Laplacian $\del^{-2}$ repeatedly $m$ times.  In Fourier space it is diagonalized, and its definition may therefore be extended to fractional values for $m$. The same holds for $k$. 

The case $\gk = 1$ corresponds to  standard molecular viscosity.  The interaction of the atmosphere with the viscous Ekman boundary layer introduces an  energy sink to the interior fluid, known as Ekman damping, that corresponds to the case $m=0$ \cite{book:Pedlosky:1979}. The same case also seems to describe an energy dissipation mechanism in soap film experiments \cite{article:Wu:2000}. In this sense, one may claim that the case $m=0$ is ``physical'' and the case $m>0$ is ``artificial'', or numerical. 

\subsection{Reformulations of Governing Equations}

To eliminate pressure we multiply both sides of the Navier-Stokes equation with the operator $\cP_{\ga\gb} \equiv \gd_{\ga\gb} - \pda\pdb\ilapl$ and we employ $\cP_{\ga\gb} u_{\gb} = u_{\gb}$ and $\cP_{\ga\gb} \pdb = 0$ to obtain 
\begin{equation}
\pderiv{u_{\ga}}{t} + \cP_{\ga\gb}\pdc (u_{\gb} u_{\gc}) = \cD u_{\ga} + \cP_{\ga\gb} f_{\gb}.
\end{equation}
 The operator $\cP_{\ga\gb}$ can be expressed in terms of a kernel  $ P_{\ga\gb} (\bfx)$ as 
\begin{align}
\cP_{\ga\gb} v_{\gb} (\bfx) &= \int d\bfy P_{\ga\gb} (\bfx-\bfy)  v_{\gb} (\bfy) \\
 &= \int d\bfy P_{\ga\gb} (\bfy)  v_{\gb} (\bfx-\bfy).
\end{align}
 For two-dimensional turbulence  $ P_{\ga\gb} (\bfx)$ is given by 
 \begin{equation}
 P_{\ga\gb} (\bfx)= \gd_{\ga\gb} \gd (\bfx) - \frac{1}{2\pi}\left[ \frac{\gd_{\ga\gb}}{r^2} - 2 \frac{x_{\ga}x_{\gb}}{r^4} \right].
\end{equation}

The scalar vorticity  $\gz$ is given by $\gz = \gee_{\ga\gb} \pd_\ga u_\gb$ with $\gee_{\ga\gb}$ the Levi-Civita tensor in two dimensions. From the incompressibility condition $\pd_\ga u_\ga = 0$ it follows that there is a function $\gy$, called the streamfunction, such that $u_\ga = \gee_{\ga\gb}\pd_\gb \gy$. Using the identity $\gee _{\ga\gb}\gee _{\gb\gc} = \gd_{\ga\gc}$ one then shows that $\gz = \gee_{\ga\gb}\gee_{\gb\gc}\pd_\ga\pd_\gc \gy = \del^2 \gy$ from which we get $\gy = \del^{-2} \gz$ and $u_\ga = \gee_{\ga\gb} \pd_\gb \del^{-2} \gz$.

The vorticity equation is obtained by differentiating $\gz$ with respect to time and employing the Navier-Stokes equations:
\begin{equation}
\pderiv{\gz}{t} + J(\gy, \gz) = \cD \gz + g, \label{eq:vorticity}
\end{equation}
where $J(\gy, \gz)$ is the Jacobian defined as
\begin{equation}
J(A, B) = \gee_{\ga\gb} (\pd_\gb A)(\pd_\ga B),
\end{equation}
and $g = \gee_{\ga\gb} \pd_\ga f_\gb$ is the forcing term. The nonlinear term $J \equiv J (\gy, \gz)$ has been obtained by employing the following argument
\begin{align}
J &= \gee_{\ga\gb}\pd_\ga \cP_{\gb\gc}\pd_{\gd} (u_\gc u_\gd ) = \gee_{\ga\gb}\pd_\ga [u_\gc \pd_\gc u_\gb] \\
&= u_\gc \pd_\gc \gz + (\gee_{\ga\gb}\pd_\ga u_\gc) (\pd_\gc u_\gb) \\
&= u_\gc \pd_\gc \gz = J(\gy,\gz).
\end{align}
The term $(\gee_{\ga\gb}\pd_\ga u_\gc) (\pd_\gc u_\gb)$ represents vortex stretching, but in two dimensions it can be shown that 
\begin{equation}
(\gee_{\ga\gb}\pd_\ga u_\gc) (\pd_\gc u_\gb) =0.
\label{eq:notilting}
\end{equation}
 by direct substitution of the vector components.

\subsection{Conservation Laws}

 The critical feature that distinguishes two-dimensional turbulence from three-dimensional turbulence is that there are two relevant conservation laws rather than just one.

It can be shown that  if two arbitrary fields $a (\bfx, t)$ and $b (\bfx, t)$ satisfy a homogeneous (Dirichlet or Neumann) boundary condition, then $\snrm{J(a,b)} = 0$, where we use the notation  
\begin{equation} 
\snrm{f}\equiv \iint \avg{f(x,y)}\; dxdy,
\end{equation}
for the combined spatial and ensemble average. It follows from the product rule of differentiation that
\begin{equation}
\snrm{J(ab,c)} = \snrm{aJ(b,c)} + \snrm{bJ(a,c)} = 0,
\end{equation}
from which we obtain the identity
\begin{equation}
\snrm{aJ(b,c)} = \snrm{bJ(c,a)} = \snrm{cJ(a,b)},
\end{equation}
which was also shown previously by Tran and Shepherd \cite{submitted:KLB}. This identity can be used to derive the conservation laws. 

The equation  $\pderivin{\gz}{t} + J(\gy,\gz) = 0$ conserves the enstrophy  $G = (1/2)\snrm{\gz^2}$  because
\begin{equation}
\snrm{\dot G} = \snrm{\gz\dot\gz} = \snrm{-\gz J(\gy,\gz)} = \snrm{-\gy J(\gz, \gz)} = 0.
\end{equation}
To derive the conservation of energy we first note that the Laplacian operator $\lapl$ and the inverse Laplacian operator $\ilapl$  are both  self-adjoint in the sense that they satisfy $\snrm{f (\lapl g)} = \snrm{g (\lapl f)}$ and  $\snrm{f (\ilapl g)} = \snrm{g (\ilapl f)}$for any fields $f(x,y)$ and $g(x,y)$. The self-adjoint property of the  inverse Laplacian operator $\ilapl$ implies that  the energy  $E \equiv (1/2)\snrm{-\gy\gz}$ is also conserved via the following argument:
\begin{align}
\snrm{\dot E} &= (1/2)\snrm{-\gy\dot\gz - \gz\dot\gy} \\
&= (1/2)[\snrm{\gy J(\gy,\gz)} +  \snrm{\gz\ilapl J(\gy,\gz)}] \\
&=  (1/2)[\snrm{\gy J(\gy,\gz)} + \snrm{\ilapl\gz)J(\gy,\gz)}] \\
&=  \snrm{\gy J(\gy,\gz)}  = \snrm{\gz J(\gy,\gy)} = 0.
\end{align}

The  energy spectrum $E (k)$ and  the enstrophy spectrum  $G (k)$ are defined as
\begin{align}
E (k) &= \frac{1}{2} \frac{d}{dk}\snrm{-\gy^{<k}\gz^{<k}}, \\
G (k) &= \frac{1}{2} \frac{d}{dk}\snrm{(\gz^{<k})^2},
\end{align}
with $\gy^{<k}$ and $\gz^{<k}$ the streamfunction and vorticity fields
with all the Fourier wavenumbers greater than $k$ in magnitude
filtered out. The spectral equations are obtained by differentiating $E(k)$ and $G(k)$ with respect to $t$, and employing the Fourier transform of the governing equation \eqref{eq:vorticity}: 
\begin{align}
\frac{\partial E (k)}{\partial t}  + \frac{\partial \Pi_E (k)}{\partial k}  &=
-D_E (k) + F_E (k)\label{eq:consE}\\
\frac{\partial G (k)}{\partial t}  + \frac{\partial \Pi_G (k)}{\partial k}  &=
-D_G (k) + F_G (k)\label{eq:consG}.
\end{align}
 It is understood that ensemble averages have been taken in the above quantities. Here $\Pi_E (k)$ is the energy flux  transfered from $(0,k)$ to $(k,+\infty)$ per unit time by the nonlinear term in \eqref{eq:vorticity}, $D_E (k)$ the  energy dissipation, and $F_E (k)$ the energy  forcing spectrum, and likewise for the enstrophy ($G$) equation. The conservation laws imply for the viscous case that $\Pi_E (0) = \lim_{k\goto\infty} \Pi_E (k) = 0$ and  $\Pi_G (0) = \lim_{k\goto\infty} \Pi_G (k) = 0$. For the inviscid case, this condition can be violated, in principle, by anomalous dissipation for solutions that have singularities. The energy and enstrophy spectrum  are related as $G(k)= k^2 E(k)$, and likewise it is easy to show that $D_G(k)= k^2 D_E (k)$ and  $F_G (k)=k^2 F_E(k)$. Combining these equations with \eqref{eq:consE} and \eqref{eq:consG} we obtain the so-called Leith constraint \cite{article:Leith:1968}:
\begin{equation}
\pderiv{\Pi_G (k)}{k} = k^2  \pderiv{\Pi_E (k)}{k}.
\end{equation}

\subsection{Direction of Fluxes}

It was recognized by  \Fjortoft \cite{article:Fjortoft:1953} and Charney \cite{article:Charney:1971}  that the direction of net energy transfer in 2D and QG turbulence may be different from that for 3D isotropic and homogeneous turbulence and that the cause for this different behavior should be attributed to the former's twin conservation of energy and enstrophy. It is often  claimed that \Fjortoft has shown that if a unit of energy is moved downscale, many more units of it have to be moved upscale in order to preserve the twin energy and enstrophy conservation. However, \Fjortofts analysis of triadic transfers   was  flawed \cite{article:Warn:1975,article:Welch:2001,submitted:Gkioulekas:2}.  His proof made use of the simultaneous conservation of energy and enstrophy in 2D and QG turbulence, and the fact that enstrophy spectrum $G(k)$ is related to the energy spectrum $E(k)$  by $G(k) = k^2 E(k)$. 

In his paper, \Fjortoft \cite{article:Fjortoft:1953} gives two distinct proofs. The first proof  does show that the only admissible triad interactions  are those that spread energy from the middle wavenumber to the outer wavenumbers (and vice versa, the ones that bring in energy to the central wavenumber from the outer wavenumbers). These are the triad interactions defined by Waleffe \cite{article:Waleffe:1992} as class ``R''. An alternative set of triad interactions are the ones where  energy is transfered from the smallest wavenumber to the two largest ones; these are the class ``F'' triad interactions, and they are dominant in three-dimensional turbulence. Fj{\o}rt{\o}ft's proof can be employed to rule these out in two-dimensional turbulence. \emph{However}, as was pointed out by  Merilees and Warn \cite{article:Warn:1975} there exist also class ``R'' triad interactions that transfer more energy downscale  than upscale. Thus, eliminating  the class ``F'' interactions is not sufficient to constrain the direction of the energy flux or the enstrophy flux.

To see this more clearly, let us carefully reconsider \Fjortofts \cite{article:Fjortoft:1953} argument, using the notation of Kraichnan \cite{article:Kraichnan:1967:1}. Let $T(k,p,q)$ be the energy transfer to the wavenumber shell $k$ from the wavenumber shells $p$ and $q$. Detailed conservation of energy and enstrophy implies that
\begin{align}
T(k,p,q) &+ T(p,q,k) + T(q,k,p) = 0 \\
k^2 T(k,p,q) &+ p^2 T(p,q,k) + q^2 T(q,k,p) = 0.
\end{align}
Solving for $T(p,q,k) $ and $T(q,k,p) $ in terms of $T(k,p,q)$ we find:
\begin{equation}
T(p,q,k) = \frac{k^2-q^2}{q^2-p^2}\; T(k,p,q)
\quad \text{and} \quad
T(q,k,p) = \frac{p^2-k^2}{q^2-p^2}\; T(k,p,q).
\end{equation}
Let us assume that $p<k<q$. We see that $T(k,p,q)<0$ implies that $T(p,q,k)>0$ and $T(q,k,p)>0$. Thus, class ``F'' triad interactions can be ruled out. The transfer difference towards the ``outer'' wavenumbers $p$ and $q$ from the inner wavenumber $k$ is given by:
\begin{equation}
\gD T(k,p,q) \equiv T(p,q,k)-T(q,k,p) =  \frac{2k^2-p^2-q^2}{q^2-p^2}\; T(k,p,q).
\end{equation}
For the traditional example $p=k/2$ and $q=2k$, we have $\gD T(k,p,q) = -(9/15)T(k,p,q)$ which is in the upscale direction when $T(k,p,q) <0$ (i.e.  when energy goes to the outer wavenumbers). However, for $q^2 = \gl (2k^2 - p^2)$ we have
\begin{equation}
\gD T(k,p,q) = \frac{(2k^2-p^2)(1-\gl)}{q^2-p^2}\; T(k,p,q).
\end{equation}
We see that the transfer difference is upscale only when $\gl > 1$ and is in fact downscale when $\gl <1$. The constraint $q>k$ implies $\gl > k^2/(2k^2 - p^2)$, however it is easy to show that $p<k$ implies $1>k^2/(2k^2 - p^2)$. So there is a critical region 
\begin{equation}
\frac{k^2}{2k^2 - p^2}<\gl <1,
\end{equation}
for the parameter $\gl$ where the outgoing triad interactions transfer more energy downscale than upscale for all $p<k$. Despite this problem, Fj{\o}rt{\o}ft's proof has been popularized in textbooks \cite{book:Salmon:1998} and review articles \cite{article:Tabeling:2002} as a rigorous proof that constrains the direction of the fluxes in two-dimensional turbulence, thereby becoming a bit of a misunderstood ``folklore'' argument. 

The second result of \Fjortoft \cite{article:Fjortoft:1953} is an upper bound on the total energy accumulated on wavenumbers larger than some given $k$. This result however applies only to initial value problems without forcing, where energy has to be bounded, unsurprisingly.  This inequality was later taken by Charney \cite{article:Charney:1971} as a proof that energy cannot go downscale, since the  energy $E^{>k}(t)$ accumulated at wavenumbers larger than $k$ is bounded by 
\begin{align}
E^{>k}(t) &= \int_{k}^{\infty} E(q)\; dq \leq \frac{1}{k^2} \int_{k}^{\infty} q^2 E(q)\; dq \leq
\frac{1}{k^2} G (t) \leq \frac{1}{k^2}G(0),
\end{align}
where $G(t)$ is the total enstrophy at time $t$. Thus, the energy spectrum $E(k,t)$ is bounded by $E(k,t) \leq c k^{-3}$ for some constant $c$. Tung and Welch \cite{article:Welch:2001} pointed out that this behavior of the energy spectrum is merely a consequence of the requirement for convergence of the Fourier representation of the enstrophy spectrum $G(k)$, which implies that $G(k)$ must decay faster than $k^{-1}$ as $k \goto \infty$.  Therefore the energy spectrum $E(k)$ must decay faster than $k^{-3}$ as $k \goto \infty$.  It says nothing about the direction of energy cascade, thus it does not help Fj{\o}rt{\o}ft's ``proof'' in the first half of the paper.

Other proofs \cite{article:Rhines:1975,article:Rhines:1979,book:Salmon:1998,article:Scott:2001,article:Eyink:1996} have been reviewed recently by Gkioulekas and Tung \cite{submitted:Gkioulekas:2}. In that paper \cite{submitted:Gkioulekas:2} we have also given a general   unified proof both for the forced-dissipatice case and for the decaying case: Assume that the forcing spectrum $F_E (k)$ is confined to a narrow interval of wavenumbers $[k_1, k_2]$. Then, we have
\begin{equation}
F_E (k) = 0 \text{ and } F_G (k) = 0, \forall k\in (0, k_1)\cup (k_2, +\infty), \label{eq:forceassumption}
\end{equation}
and it can be shown \cite{submitted:Gkioulekas:2} for the forced dissipative case, without making any ad hoc assumptions, that   under stationarity, the fluxes $\Pi_E (k)$ and $\Pi_G (k)$ will satisfy the inequalities
\begin{align}
\int_0^k  q \Pi_E (q) \; dq &<0,  \;\forall k > k_2 \label{eq:ineqE}\\
\int_k^{+\infty} q^{-3} \Pi_G (q) &> 0, \;\forall k < k_1 \label{eq:ineqG}.
\end{align}
The constraint \eqref{eq:ineqE} holds trivially for $k<k_1$,  since $\Pi_E (k) < 0$ for all $k<k_1$.  For $k>k_2$, the integration range also includes the energy injection interval $[k_1, k_2]$ and both the upscale cascade range and the downscale cascade range.  The inequality \eqref{eq:ineqE} implies that the negative flux in the $(0,k_1)$ interval is more intense than the positive flux in the $(k_2, +\infty)$ because the weighted average of $\Pi_E (k)$ gives more weight to the large wavenumbers. Thus, \eqref{eq:ineqE}  implies that energy fluxes upscale in the net. Similarly, \eqref{eq:ineqG} implies that enstrophy fluxes downscale in the net.  These results can be extended \cite{submitted:Gkioulekas:2} to the decaying case provided that there  exists a small wavenumber $\gee_1 >0$ and a large wavenumber $\gee_2 > 0$ such that
\begin{align}
D_A (q) &+\pderiv{A(k)}{t} \geq 0 ,\;\forall k\in(0, \gee_1)\label{eq:assE}\\
D_A (q) &+\pderiv{A(k)}{t}\geq  0 ,\;\forall k\in (\gee_2, +\infty).\label{eq:assG}
\end{align}
Note that \eqref{eq:assE} implies \eqref{eq:ineqE} and \eqref{eq:assG} implies \eqref{eq:ineqG}. 

It should be noted that, unlike previous proofs, in both the forced-dissipative and the decaying case, the inequalities \eqref{eq:ineqE} and \eqref{eq:ineqG} have the same mathematical form. Our argument then is a unified proof that covers all cases, and specialized results can be deduced from our inequalities for special cases. Note that none of these results  forbids energy from being transferred downscale even when it is shown that the net flux should be directed upscale; they merely say that in those cases the energy going upscale in the upscale range should be larger than that going downscale in the downscale range.

\section{Phenomenology of 2D Turbulence}

 Kraichnan \cite{article:Kraichnan:1967:1}, Leith
 \cite{article:Leith:1968}, and Batchelor
 \cite{article:Batchelor:1969}  (KLB) proposed that in two-dimensional
 turbulence there is an upscale energy cascade and a downscale
 enstrophy cascade,  when stochastic forcing injects energy and enstrophy in a narrow band of intermediate length scales.  This scenario was inspired by Kolmogorov's idea \cite{article:Kolmogorov:1941,article:Kolmogorov:1941:1,article:Batchelor:1947}
of an inertial range for three-dimensional turbulence within which
energy cascades from large scales by hydrodynamic instability to small
scales. Assuming an infinite domain where  all the energy flows upscale and all of the enstrophy flows downscale, KLB invoke a dimensional analysis argument, similar to Kolmogorov's, to show that the energy spectrum in the upscale energy  range is 
 \begin{equation}
E(k) = C_{ir}\gee^{2/3}k^{-5/3}, \;\text{ for } k\ell_0 \ll 1,
\end{equation}
and in the downscale enstrophy  range is 
\begin{equation}
E(k) = C_{uv}\gn^{2/3}k^{-3}, \;\text{ for } k\ell_0 \gg 1. 
\end{equation}
Anticipating the objection that the dimensional analysis arguments cannot be applied to the enstrophy cascade, because of nonlocality, in a subsequent paper \cite{article:Kraichnan:1971:2} Kraichnan  proposed that the enstrophy cascade energy spectrum is given by 
\begin{equation}
 E(k) = C_{uv}\gn^{2/3}k^{-3}[\ln (k\ell_0)]^{-1/3}, 
\end{equation}
and showed, using  a one-loop closure model \cite{article:Kraichnan:1971:1}, that this logarithmic correction is consistent with constant enstrophy flux.  The  same result can be obtained with other 1-loop models  \cite{article:Ishihira:2001}, and has been refined further by Bowman \cite{article:Bowman:1996}.

In its traditional form, the KLB scenario  incorporates two unrealistic assumptions. First, it requires an unbounded domain to allow the upscale energy flux to escape to larger and larger length scales without the need for infrared dissipation. A number of recent theoretical results \cite{article:Shepherd:2002,article:Bowman:2003,article:Bowman:2004} challenge the realizability of cascades as envisaged by Kraichnan for the standard case of Navier-Stokes \emph{without} an infrared sink in a \emph{bounded} domain. In the more realistic case of a finite domain, a dissipative sink is needed both at large scales and at small scales, in order for cascades to form. 
 Second, in the KLB scenario the energy dissipation sink at the small scales is taken to approach zero so that the dissipation scale occurs at infinite wavenumber, which is unrealistic in geophysical formulations \cite{article:Orlando:2003:1,article:Tung:2004} and in numerical simulations.  The problem we are considering is finite domain with finite dissipation both at large scales and small scales, which is more realistic in geophysical applications.

\subsection{Numerical Results}

In recent years, it has been possible to reproduce either  the direct enstrophy cascade \cite{article:Alvelius:2000,article:Kaneda:2001,article:Falkovich:2002} or the inverse energy cascade \cite{article:Aref:1981,article:Sulem:1984,article:McWilliams:1985,article:Tabeling:1997,article:Tabeling:1998,article:Vergassola:2000}, but not both simultaneously, in numerical simulations.  Even these have  some interesting complications, which are briefly discussed bellow. 

 For the case of the downscale enstrophy cascade, it has been found that the presence of coherent structures at large scales can prevent its development.  Borue  \cite{article:Borue:1993} showed that using hypodiffusion ($m=8$ and $k=1,8$) disrupts the coherent structures and with increasing Reynolds number the scaling of the enstrophy range approaches asymptotically Kraichnan scaling.  As pointed out by Tran and Shepherd \cite{article:Shepherd:2002}, all the successful simulations of the $k^{-3}$ spectral range are done so far with the hypodiffusion device. As a matter of fact, Bernard \cite{article:Bernard:2000} has given an elementary proof that under Ekman damping it is not possible for the energy spectrum of the downscale cascade to scale precisely as $k^{-3}$ with or without the logarithmic correction. A steeper energy spectrum is predicted instead. Nam \emph{et al.} \cite{article:Ott:1999}  have derived a law governing the steepening of the enstrophy cascade by Ekman damping, however it cannot be used directly to predict the slope of the energy spectrum from the viscosity parameters without additional experimental input. On the other hand, there is sufficient recent  evidence from numerical simulations to show that under hypodiffusion, an enstrophy cascade, consistent with the KLB theory, can be obtained if the numerical resolution is sufficiently large \cite{article:Alvelius:2000,article:Falkovich:2002,article:Ishihira:2001}. This was not the case even a few years earlier, when various spectral slopes steeper than $-3$ were found in numerical simulations prevailing at the time.

A similar problem arises for the case of the inverse energy cascade.  Danilov and Gurarie \cite{article:Gurarie:2001}  have conducted numerical simulations using $(m,\gk)=(0,2)$, and showed that the optimal $\gb$ yielding an  energy spectrum closest to the KLB prediction of $k^{-5/3}$ scaling does not correspond to constant energy flux.   Decreasing $\gb$ improves the energy flux but the slope of the energy spectrum steepens.  This behavior is somewhat minimized in simulations using $(m,\gk)=(0,8)$, but the reverse relation between optimizing the flux and optimizing the spectrum persists. Sukoriansky \emph{et al } \cite{article:Chekhlov:1999} noted that using higher order large-scale hypo-dissipation ($m>0$) may produce a constant energy flux, but distorts the spectrum. It has therefore been suggested that the locality of the inverse energy cascade should be called into question \cite{article:Danilov:2003}.  Both Danilov and Gurarie \cite{article:Gurarie:2001:1}, and  earlier Borue \cite{article:Borue:1994}, observed that this steepening is caused by coherent structures.  These structures cover a relatively small portion of the domain, but they account for most of the energy.   

The most convincing evidence that we have in support of the existence of the inverse energy cascade is the numerical simulation by Boffetta \emph{et al} \cite{article:Vergassola:2000}.  They used $2048^2$  resolution and Ekman damping ($m=0$), and obtained not only the $k^{-5/3}$ energy spectrum, but also the $3/2$-law, which is the mathematical equivalent of constant energy flux in real space. Ironically, the energy flux in Fourier space is still not constant.  So far as we know, an inverse energy cascade with $k^{-5/3}$ energy spectrum and constant energy flux in Fourier space hasn't been produced by any of the numerical simulations reported in the literature.

  Boffetta \emph{et al} \cite{article:Vergassola:2000} are aware of the steepening of the energy spectrum in the numerical simulation of Borue \cite{article:Borue:1994} where hypodiffusion is used instead of Ekman damping, and explain it in terms of the ``bottleneck effect''.  This effect is essentially a distortion of the solution corresponding to the inverse energy cascade by the dissipation operator acting at large scales.  It has been observed in the direct energy cascade of three-dimensional turbulence \cite{article:Falkovich:1994:1} and earlier in acoustic turbulence \cite{article:Ryzhenkova:1990}.  While we do not question this possibility, another possible culprit is the violation of homogeneity by the existence of periodic boundary conditions at large scales.  The locality of the inverse energy cascade relies on the elimination of the sweeping interactions of the large-scale shear flow; this violation of homogeneity can cause the sweeping interactions to have a significant nonlocal effect on the energy spectrum of the inverse energy cascade \cite{article:Gkioulekas}, which may be the theoretical origin of the well-known condensate.  Ironically, it has been shown that in a downscale enstrophy cascade dominated by  sweeping, the energy spectrum is still $k^{-3}$ but without the logarithmic correction \cite{article:Laval:2000,article:Nazarenko:2000}.

The conclusion drawn from the numerical work so far is that while both cascades can manifest themselves under favorable conditions, neither cascade is completely robust.  This should be contrasted with the situation in three-dimensional turbulence where the energy cascade is easily reproduced in numerical simulations, both forced and decaying.  It is ironic that it has proven more difficult to produce the cascades of two-dimensional turbulence in numerical simulations in light of the advantage of working with only two dimensions.


\subsection{Theoretical Results}

Because numerical simulations did not reproduce the $k^{-3}$ energy spectrum of  the downscale range consistently, alternative theories have been proposed that predict steeper scaling \cite{article:Saffman:1971,article:Polyakov:1993,article:Moffatt:1986}.  Kraichnan \cite{article:Kraichnan:1967:1} himself noted that the  non-locality of the direct enstrophy cascade makes the application  of dimensional analysis inconsistent, unless a logarithmic correction is introduced, as explained earlier.    If higher order closures yield additional higher powers of   logarithmic corrections, they could  add up to a power law   renormalization leading to a steeper spectrum. 

Eyink has shown   recently  \cite{article:Eyink:2001} that such a renormalization does   not take place when a downscale enstrophy cascade manifests with   constant enstrophy flux, although logarithmic corrections are not   excluded. This result is based on the mathematical theory of DiPerna   and Lions \cite{article:Lions:1989}, and it is a refinement of an earlier argument \cite{article:Eyink:1995,article:Eyink:1996} that only ruled out energy spectra steeper than  $k^{-11/3}$. The main result of Eyink \cite{article:Eyink:2001} is that in the inviscid limit  $\nu\goto 0$ there is no anomalous enstrophy sink if the total enstrophy remains finite.  It follows that the total enstrophy must diverge when $\nu\goto 0$, otherwise a downscale enstrophy cascade is not possible.  This rules out any energy spectrum that is steeper than $k^{-3}$, although the scaling $k^{-3}[\ln (k\ell_0)]^{-p}$ with $0\leq p <1$ , is allowed \cite{comm:Eyink}.  

The absence of intermittency corrections in the downscale enstrophy cascade can be  understood conceptually also by comparing against the case of the downscale energy cascade in three-dimensional turbulence.  It is well known that in three-dimensional turbulence intermittency corrections arise from logarithmic contributions of ladder-type Feynman diagrams to the response functions  in the fusion limit, which bring out  the same scaling exponent as the usual structure functions \cite{lect:Procaccia:1994,article:Procaccia:1995:1,article:Procaccia:1995:2,article:Procaccia:1996,article:Procaccia:2000}.  These logarithmic contributions have a numerical coefficient $\gD = 2-\gz_2$.  For the case of the enstrophy cascade, $\gz_2=2$ implies that the coefficient $\gD$ is zero, so the  mechanism which causes   intermittency corrections in three-dimensional turbulence, is lost in two-dimensional turbulence.

  Falkovich and Lebedev \cite{article:Lebedev:1994,article:Lebedev:1994:1} used a Lagrangian approach \cite{article:Lebedev:1995:1,article:Vergassola:2001} to confirm Kraichnan scaling with the logarithmic correction. They also predict that the vorticity structure functions have regular (i.e. the scaling exponents form a straight line) logarithmic scaling given by  
\begin{equation}
\left\langle [\gz (\bfr_1) - \gz (\bfr_2)]^n \right\rangle \sim [\gn\ln (\ell_0/r_{12})]^{2n/3}.
\end{equation}
Eyink \cite{article:Eyink:1996:1} noted that this theory does not follow from first principles and that it rests on an unproven regularity for the velocity field.  However, should this regularity condition be proven, it would then follow that the Kraichnan scaling scenario is the only one that is statistically stable \cite{article:Lebedev:1996}. 

Although Eyink's most recent result \cite{article:Eyink:2001} shows that intermittency corrections are excluded in the enstrophy cascade, the argument still relies on the successful formation of an enstrophy cascade under given configurations of dissipation at large scales and small scales.  An additional argument is then needed to show that  the enstrophy cascade forms  successfully and is in some sense local and universal.  Together with this argument, the results of Eyink \cite{article:Eyink:2001} combined with the theory by Falkovich and Lebedev \cite{article:Lebedev:1994,article:Lebedev:1994:1} give a satisfying theory for the downscale enstrophy cascade.

Unlike the enstrophy range, Kraichnan's dimensional analysis argument is at least self-consistent for the inverse energy cascade.  A recent numerical simulation \cite{article:Vergassola:2000} has revealed that there are no intermittency corrections to the inverse energy cascade.  This has also been corroborated by experiment \cite{article:Tabeling:1998}. Yakhot  \cite{article:Yakhot:1999} has formulated an interesting theoretical explanation, using a mathematical technique developed by Polyakov \cite{article:Polyakov:1995}  for Burger's turbulence.  His argument is based on the  assumption that the pressure gradients are local, in the sense of his theoretical framework. The physical meaning of this assumption, in our opinion,  is that there \emph{does} exist  an inverse energy cascade, where local interactions are dominant, without any trouble from the bottleneck effect, sweeping interactions, or (theoretically possible but unlikely) instability with respect to the forcing at small scales.

A recent paper by L'vov \emph{et al} \cite{article:Procaccia:2002} provides further insight into the nature of the inverse energy cascade.  It is shown that for hydrodynamic turbulence with dimension $d=4/3$, a $k^{-5/3}$ energy spectrum at large scales is an \emph{enstrophy absolute equilibrium spectrum} where the velocity field is completely Gaussian and the energy does not flow in either direction.  When the dimension is increased to the physical value $d=2$, the velocity field remains mostly Gaussian but an upscale energy flux is now allowed.  As a result, intermittency corrections are sufficiently negligible that they're not observable. 

A comprehensive theory of the two-dimensional inverse energy cascade still remains an unfinished task.

\section{The Nastrom-Gage Spectrum}

According to Kraichnan \cite{article:Kraichnan:1967:1}, the study of two-dimensional turbulence was motivated by the hope that it would prove a useful model for atmospheric turbulence. This idea was later encouraged  by Charney \cite{article:Charney:1971} who claimed that quasi-geostrophic turbulence is isomorphic to two-dimensional turbulence. The question that was then posed was whether the energy spectrum of the atmosphere at length scales that are orders of magnitude larger than the thickness of the atmosphere can be explained in terms of the theory of two-dimensional turbulence. This question continues to be debated today.

 Early observations by Wiin-Nielsen \cite{article:Wiin-Nielsen:1967} suggested that the energy spectrum of the atmosphere follows a $k^{-3}$  power law behavior consistent with an enstrophy cascade. Because of the  sparseness of observational stations, only results for the planetary scales ($\sim 10000$km) and synoptic scales ($\sim 1000$km) were shown. Wiin-Nielsen's data at the time appeared to fit this picture, with approximately a $-3$ power law for wavenumbers between 8 and 16, and a (less defined) $-0.4$ power law for wavenumbers smaller than 8.  The break in the slopes was identified \cite{article:Wiin-Nielsen:1967,article:Wiin-Nielsen:1978} as the location of energy injection by baroclinic instability, which was assumed to occur in a narrow wavenumber band around 8.

\begin{figure}
\begin{center}
\scalebox{0.6}{\includegraphics{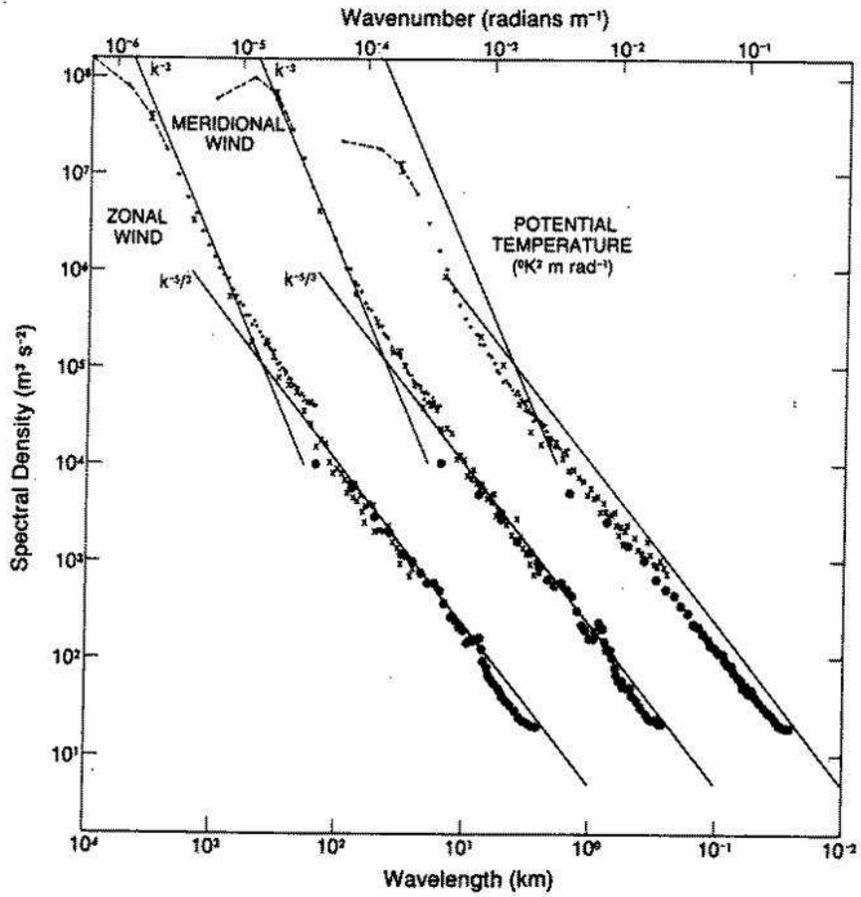}}
\caption[]{\small The Nastrom-Gage spectrum of atmospheric turbulence. For the purpose of showing the individual spectrum, the spectrum for meridional wind is shifted one decade to the right, while that for potential temperature is shifted two decades to the right. }
\label{fig:NG}
\end{center}
\end{figure}

An analysis   by Nastrom and Gage  \cite{article:Jasperson:1984,article:Gage:1984} of high resolution wind and temperature measurements, collected using  commercial airplane flights in the upper troposphere and lower stratosphere in the late 70's,   showed that there is a robust $k^{-3}$ spectrum extending from approximately 3000 km to 800 km in wavelength (the ``synoptic scales'') and a robust $k^{-5/3}$ spectrum extending from 600 km down to a few kilometers (the ``mesoscales''). The transition from one slope to the other occurs gradually between $600$km and $800$km.  Recent measurements \cite{article:Cook:1998,article:Albercook:1999,article:Barrick:1990} have confirmed the $k^{-5/3}$ part of the atmospheric energy spectrum. This remarkably robust spectrum is widely known as the Nastrom-Gage spectrum (see Figure~\ref{fig:NG}). General Circulation  models have been shown to be capable of reproducing the Nastrom-Gage spectrum in agreement with observations \cite{article:Mahlman:1999,article:Hamilton:2001,article:Skamarock:2004}.

\subsection{Prior Explanations of the Nastrom-Gage Spectrum}

Naively one could interpret the $-3$ slope as two-dimensional (2D) turbulence, for which Kraichnan  predicted the $-3$ slope for downscale enstrophy  cascade. One could also interpret the $-5/3$ sloped portion of the spectrum for scales less than 600 km as three-dimensional (3D) turbulence, for which Kolmogorov predicted a $-5/3$ spectral slope for homogeneous isotropic 3D turbulence with downscale energy cascade, and attribute the transition from the $-3$ slope to the $-5/3$ slope as due to the motion being predominantly 2D for the large scales in the thin shell of the troposphere and the shorter scales as being 3D.   However, results of classical 3D turbulence cannot be applied here to motions whose horizontal scales range up to 600km in wavelength, while its vertical scale is about 10km (which is the scale of the depth of the troposphere).  The $-3$ sloped portion could conceivably be explained by Quasi-Geostrophic (QG) turbulence of Charney \cite{article:Charney:1971}, which bears much resemblance to 2D turbulence. However  caution should be exercised over most of the scales involved (the synoptic scales), which are in the forcing range where energy and enstrophy injection occurs and therefore are not in an inertial range \cite{article:Tung:1998,article:Welch:2001}.

Some of the earliest theories for the $-5/3$ part of the spectrum involved internal gravity waves \cite{article:Dewan:1979,article:VanZadt:1982}.   As pointed out by Gage and Nastrom \cite{article:Nastrom:1986}, observational information on the vertical velocity spectrum revealed that there is a basic inconsistency between the observed spectra and theories of internal waves as the cause of the mesoscale spectrum.  There is ``simply too much energy'' in the horizontal spectrum compared to the vertical spectrum to be consistent with the idea that both are due to a common spectrum of internal waves.

A proposed theory still being considered is due to Lilly \cite{article:Lilly:1989}, who suggested that thunderstorms at the short-wave end (at km scales) of the spectrum have enough energy to generate stratified turbulence which then collapses into 2D turbulence.  This 2D turbulence cascades upscale to form the $-5/3$ portion of the 2D spectrum due to the negative energy flux, in the same way as positive energy flux in  Kolmorov's theory for 3D turbulence generate its $-5/3$ spectrum (since the dimensional argument for the $-5/3$ spectral slope is independent of the sign of the energy flux). According to this theory, in addition to the small-scale source there is also a large-scale source injecting potential enstrophy and thereby giving rise to the $k^{-3}$ spectrum at large scales.  A variation on this theme is the theory of Falkovich \cite{article:Falkovich:1992:1} where the $k^{-3}$ portion is explained as a condensate and not as an enstrophy cascade.

How much of the thunderstorm's energy, which consists of gravity waves, which radiate away, and in the form of 3D turbulence, which naturally tends to cascade into still smaller scales, can be converted into 2D turbulence and cascaded up three decades of scales is questionable.  Extensive numerical calculations of stratified turbulence show that only 2\% of the energy is converted into 2D turbulence \cite{article:Waleffe:1996}.  That may or may not be sufficient to generate the observed spectrum without further study.  What is more difficult to explain with this theory of thunderstorm source of energy is the fact that the spectrum, in particular the transition wavenumber between the shallow portion and the steeper portion of the spectrum, appears to be approximately the same whether it is in winter or summer, and whether the airplane flew over storms or not.  

 At first Lilly's theory of small-scale and large-scale source was favored because Lindborg \cite{article:Lindborg:1999}, using third order structure functions, appeared to have deduced from observed data that the energy flux is upscale.  Later, however, Cho and Linborg \cite{article:Lindborg:2001,article:Lindborg:2001:1} corrected a sign error which then led them to conclude that their analyses of data at ``mesoscales in both the upper troposphere and lower stratosphere provide no support for an inverse energy cascade 2D turbulence.'' There is now even analysis \cite{article:Thornhill:2003} of aircraft data which gives the magnitude of the finite dissipation rate at the small scales (or equivalently, the downscale energy flux).

\begin{figure}
\begin{center}
\scalebox{0.6}{\includegraphics{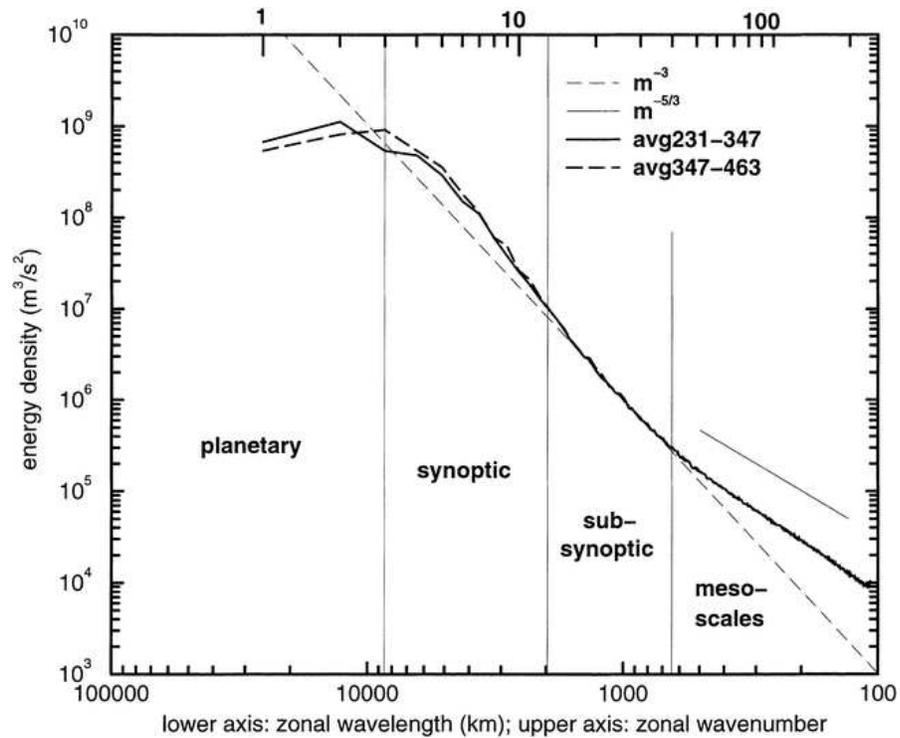}}
\caption[]{\small The energy spectrum predicted by the model of Tung and Orlando \cite{article:Orlando:2003}. Note the -3 sloped spectrum over the subsynoptic scales transitioning at around 700km to a -5/3 sloped spectrum over the mesoscales.  These features compare very favorably to the observed spectrum.  For larger scales, the synoptic and planetary scales, there were not enough long distanced flight segments in the Nastrom-Gage data, hence the drop off in power.  Nevertheless, station data can be used to supplement the aircraft data for the larger scales.  Our model results compare favorably at those scales to these data as well, including the steeper than -3 slope over the synoptic scales, which are located in the forcing region \cite{article:Tung:1998}. }
\label{fig:TO}
\end{center}
\end{figure}

\subsection{The Tung-Orlando Double Cascade Theory}

The ``remarkable degree of universality'' (Gage and Nastrom 1986) in spectral amplitude and in spectral shape over the entire range of wavelengths encompassing both the $-3$ and the $-5/3$ parts of the spectrum is ``hard to explain'' if it were due to forcing on two ends of the spectrum by two unrelated physical processes.  The spectrum is perhaps best explained by a source only at the large scales being responsible for both parts of the spectrum, as first proposed by Tung and Orlando \cite{article:Orlando:2003}.  There is a natural source for this energy injection and it is of sufficient amplitude: baroclinic instability caused by the north-south temperature of the lower atmosphere (ultimately due to the sun heating the tropics more than the high latitudes, making the atmosphere potentially unstable).  Such injection of energy and potential enstrophy occurs at the synoptic scales ($10000$km to $2000$km). 

If $\gn_{uv}$ is the downscale enstrophy flux and $\gee_{uv}$ is the downscale energy flux,  it was suggested by Tung and Orlando \cite{article:Orlando:2003} that they would coexist  on the downscale side of injection in the same inertial range and  their separate contributions to the energy spectrum would give the latter a compound spectral shape, with a $-3$ slope transitioning to a shallower $-5/3$ slope as the wavenumber increases. So, in a sense, the Tung-Orlando theory is that the entire Nastrom-Gage spectrum can be conceptualized as a double cascade of both energy and potential enstrophy.

Using a dimensional argument, the transition from $-3$ slope to $-5/3$ slope is expected to occur at the transition wavenumber $k_t$ with order of magnitude estimated by $k_t \approx \sqrt{\gn_{uv}/\gee_{uv}}$. Recent measurements and data analysis \cite{article:Lindborg:2001}  estimate $\gn_{uv} \approx 2\times 10^{-15} \text{s}^{-3}$ and $\gee_{uv} \approx 6\times 10^{-11}  \text{km}^2  \text{s}^{-3}$. From these estimates we find the mean value of the transition scale $k_t = \sqrt{\gn_{uv}/\gee_{uv}} \approx 0.57\times 10^{-2}  \text{km}^{-1}$ and $\gl_t = 2\pi/k_t \approx 1\times 10^3  \text{km}$ which has the correct order of magnitude. 

 Tung and Orlando \cite{article:Orlando:2003} have  demonstrated numerically that a two-layer quasi-geostrophic channel model with thermal forcing, Ekman damping, and hyperdiffusion  can  reproduce this  compound spectrum (see Figure~\ref{fig:TO}). The resolution of these simulations goes down to 100km in wavelength.  The diagnostic shown in figure 7 of Tung and Orlando \cite{article:Orlando:2003},  shows both the constant downscale energy and enstrophy fluxes coexisting in the same inertial range. Furthermore, Tung and Orlando \cite{article:Orlando:2003} confirmed the dimensional estimate $k_t \approx \sqrt{\gn_{uv}/\gee_{uv}}$ for the transition wavenumber $k_t$.

The Tung-Orlando theory  is contrary to the widely accepted misconception in the atmospheric science community that the argument by Fjortoft \cite{article:Fjortoft:1953} forbids a downscale energy flux in two-dimensional turbulence altogether, and through the isomorphism theorem of Charney \cite{article:Charney:1971} also in quasi-geostrophic turbulence. Various  aspects of this misconception have been clarified in recent papers  \cite{article:Warn:1975,article:Welch:2001,submitted:Gkioulekas:2} and in the present paper.

 Although the nature of the nonlinear interactions which give rise to the downscale energy flux changes from quasi-geostrophic to stratified three-dimensional in the mesoscales, as far as the energy spectrum $E(k)$ is concerned it is the existence of a downscale energy flux from the largest scales ($10000$km) to the smallest scales ($1$km) which gives rise to the  $k^{-5/3}$ slope, regardless of the character of the motion.   The recent interest \cite{article:Lindborg:2005,article:Lindborg:2006} in understanding the $k^{-5/3}$ slope in terms of three-dimensional stratified turbulence is well motivated, since it is necessary to account for length scales less than 100km in wavelength  where the quasi-geostrophic assumption fails. 
It is the view of the authors that  it is equally important to understand why the quasi-geostrophic model is capable of supporting a downscale energy cascade with $k^{-5/3}$ scaling, because one also has to account for the existence of $k^{-5/3}$ scaling in the vicinity of  the transition range ($800$km to $600$km) where the quasi-geostrophic assumption is presumably valid. Estimates for the breakdown of the validity of the QG constraint range from $500$km to $100$km.

\subsection{Double Cascades in Two-Dimensional Turbulence}

In recent papers \cite{article:Tung:2005,article:Tung:2005:1}, we have suggested that the double cascade phenomenon takes place in pure 2D turbulence too, where it mighf be possible, under nonlinear dissipation,  to have a transition from $k^{-3}$ scaling to $k^{-5/3}$ scaling with increasing $k$. As has been pointed out by previous authors \cite{article:Borue:1994,article:Eyink:1996}, as long as the dissipation terms at large scales and small scales have finite viscosity coefficients and the inertial ranges exist, the downscale enstrophy flux will be accompanied by a small downscale energy flux, and the upscale energy flux will be accompanied by a small upscale enstrophy flux. Dimensional analysis arguments are premised on the assumption that these additional fluxes can be ignored, consequently the energy spectrum predictions obtained by such arguments are valid only to leading order. We have argued \cite{submitted:Gkioulekas:3} that although   subleading effects  can be  ignored  with impunity for strictly two-dimensional turbulence,  for models of quasi-geostrophic turbulence, such as the two-layer model,  the subleading contributions can be  important in the inertial range and cannot be safely ignored.

For the case of two-dimensional turbulence, we argued  \cite{article:Tung:2005,article:Tung:2005:1} that  the  subleading fluxes are associated with a subleading downscale energy  cascade and a subleading inverse enstrophy cascade that contribute  \emph{linearly} to the total energy spectrum in addition to the  dominant contributions. The two contributions are homogeneous  solutions of the underlying statistical theory, which is in fact  \emph{linear}.  Furthermore, the two  homogeneous solutions  are independent of each other, so the downscale energy cascade is independent of the downscale enstrophy flux $\gn_{uv}$ and the downscale enstrophy cascade is independent of the downscale energy flux $\gee_{uv}$. As a result, in the downscale inertial range, the total energy spectrum $E(k)$ has the following three contributions:
\begin{equation}
E(k) = E_{uv}^{(\gee)} (k) + E_{uv}^{(\gn)} (k) + E_{uv}^{(p)} (k), \;\forall k\ell_0 \gg 1,
\label{eq:energy}
\end{equation}
where $E_{uv}^{(\gee)} (k)$, $E_{uv}^{(\gn)} (k)$ are the contributions of the downscale energy and enstrophy cascade, given by
\begin{equation}
\begin{split}
E_{uv}^{(\gee)} (k) &= a_{uv}\gee_{uv}^{2/3} k^{-5/3}\cD_{uv}^{(\gee)} (k\ell_{uv}^{(\gee)}) \\
E_{uv}^{(\gn)} (k) &=  b_{uv}\gn_{uv}^{2/3} k^{-3}[\chi + \ln (k\ell_0)]^{-1/3} \cD_{uv}^{(\gn)} (k\ell_{uv}^{(\gn)}),
\end{split}
\end{equation}
with $\cD_{uv}^{(\gee)}$ and $\cD_{uv}^{(\gn)}$ describing the dissipative corrections. Here we use the logarithmic correction of Kraichnan \cite{article:Kraichnan:1971:2}, adjusted by the constant $\chi$ of Bowman \cite{article:Bowman:1996} for the contribution of the enstrophy cascade. We have also assumed, without explicit justification, that we may ignore the possibility of intermittency corrections to the subleading downscale energy cascade. For the downscale enstrophy cascade intermittency corrections have been ruled out by Eyink \cite{article:Eyink:2001}.  For the downscale energy cascade we conjecture that intermittency corrections are small for the same reasons as in three-dimensional turbulence. The scales $\ell_{uv}^{(\gee)}$, $\ell_{uv}^{(\gn)}$ are the dissipation length scales for the downscale energy and enstrophy cascade. Finally,  $E_{uv}^{(p)} (k) $ is the contribution  from the effect of forcing and the sweeping interactions and it represents a particular solution to the statistical theory. The latter can become significant via the violation of statistical homogeneity caused by the  boundary conditions \cite{article:Gkioulekas}. Thus, in the inertial range where  the effect of forcing and dissipation can be ignored, the energy spectrum will take the simple form in the downscale range:
\begin{equation}
E(k) \approx a_{uv} \gee_{uv}^{2/3}k^{-5/3} + b_{uv}\gn_{uv}^{2/3}k^{-3} [\chi + \ln (k\ell_0)]^{-1/3}.
\end{equation}
We see that the energy spectrum will take the slope of $-3$ for small $k$ , and $-5/3$ for large $k$ .  The transition from one slope to the other occurs at $k_t$ , given by $\gee_{uv} k_t^2 \sim \gn_{uv}$.

It should be emphasized that the formation of cascades observable in the energy spectrum is by no means guaranteed.  There are two prerequisites that need to be satisfied: first, the contribution of the particular solution $E_{uv}^{(p)}(k)$ has to be negligible both downscale and upscale of the injection scale, i.e.  
\begin{equation}
\begin{split}
E_{uv}^{(p)}(k) &\ll E_{uv}^{(\gee)}(k) + E_{uv}^{(\gn)}(k) , \;\forall k\ell_0 \gg 1 \\
E_{ir}^{(p)}(k) &\ll E_{ir}^{(\gee)}(k) + E_{ir}^{(\gn)}(k) , \;\forall k\ell_0 \ll 1.
\end{split}
\end{equation}
 Second, the dissipative adjustment $\cD_{uv}^{(\gn)} (k\ell_{uv}^{(\gn)})$ and $\cD_{uv}^{(\gee)} (k\ell_{uv}^{(\gee)})$ of the homogeneous solution has to be such that it does not destroy the power law scaling in the inertial range.  Furthermore, the dissipation scales $\ell_{uv}^{(\gn)}$ and $\ell_{uv}^{(\gee)}$ have to be positioned so that the incoming energy and enstrophy can be dissipated.

The idea of two cascades in the same wavenumber region has an interesting precedent in the case of three-dimensional turbulence, where there is interest in  understanding the double cascade of helicity and energy \cite{article:Chkhetiani:1996,article:Procaccia:1999,article:Eyink:2003}.  There, the situation is  more straightforward because the helicity cascade and the energy cascade reside in separate isotropic sectors of the $SO(3)$ group \cite{article:Procaccia:1999,article:Procaccia:2005}.  This makes it easier to argue in support of a superposition principle.  

In two-dimensional turbulence the situation is more interesting    because both cascades reside in the same isotropic sectors. The main argument in support of our conjecture was given in section 2 of Ref.~\onlinecite{article:Tung:2005}. Additional evidence is given in section 3 of the same paper. It should be noted that our main argument exploits the linearity of the exact statistical theory of two-dimensional turbulence (i.e. the complete infinite system of equations governing the relevant fully-unfused structure functions). Nonlinear results, such as the one that was proposed by Lilly \cite{article:Lilly:1989} and more recently by L'vov and  Nazarenko \cite{article:Nazarenko:2006}, follow from closure models instead of the exact theory. Likewise, phenomenological arguments with the eddy-turnover rate, such as the one by Kraichnan \cite{article:Kraichnan:1971:2}, are essentially coming out of a 1-loop nonlinear closure theory, and would also lead to nonlinear expressions for the energy spectrum. On the other hand, both closure models and the superposition principle give the same prediction for the transition wavenumber, and disagree only in the transition region.

\subsection{Criticisms of the Superposition Principle}

It should be noted that the superposition principle is not yet widely accepted by some researchers in the turbulence community \cite{comm:Eyink,comm:Lvov}.  A difficulty in accepting the idea of a downscale energy cascade in two-dimensional turbulence is the fact that the cascade is hidden, and the corresponding energy flux  is very small and vanishes rapidly with increasing Reynolds number.  The underlying question is to understand what should be meant by cascade for the case of finite viscosity, considering that the existence of a uniform energy flux is not sufficient to imply the existence of a local downscale energy cascade \cite{comm:Danilov,comm:Eyink}.

  The viewpoint that we have expressed in our previous work \cite{article:Tung:2005,article:Tung:2005:1,article:Gkioulekas,lfthesis:2006} is that an inclusive definition is that a cascade is a homogeneous solution to the MSR statistical theory of turbulence. One may then narrow down this definition to define ``local cascade'' and ``universal cascade'' in various ways. An inertial range exists when homogeneous solutions dominate the particular solution that arises from the forcing region and the sweeping interactions.  The existence of finite dissipation modifies the homogeneous solutions in the same sense that the natural modes of a linear oscillator become damped in the presence of friction.  It is hard to deny that in two-dimensional turbulence, for the downscale range, we have two homogeneous solutions: a solution that corresponds to the downscale enstrophy cascade and a solution that corresponds to the downscale energy cascade  \cite{article:Tung:2005,article:Tung:2005:1}.  Our argument is that, given the existence of these two solutions, the existence of a downscale energy flux implies that both solutions are present.  An alternative interpretation \cite{comm:Eyink} is to argue that the vanishing downscale energy flux is nothing more than a dissipative correction of the enstrophy solution, and that only the enstrophy solution is present.  The only way to distinguish between these two interpretations is to construct a counterexample where the downscale energy flux does not vanish and is in fact large enough to bring up an observable transition.  Such a counterexample would have to involve a nonlinear dissipation term that can selectively dissipate enstrophy without dissipating energy.


A more technical criticism of the superposition principle is that, for the downscale enstrophy cascade solution,  it is ambiguous whether  the reducible contributions of unlinked Feynman diagrams dominate the irreducible contributions of linked Feynman diagrams \cite{comm:Lvov}. If the reducible contributions dominate, then one should expect nonlinear cross-term contributions to the higher-order structure functions, in addition to the contributions anticipated by the superposition principle.  However, such cross-terms would be important only in the transition region and would not affect the location of the transition wavenumber or the asymptotic scaling away from the transition region.  Furthermore, in either case there will be no cross-terms for the second order structure functions, and consequently for the energy spectrum.

\subsection{The Difference Between 2D Turbulence and the Two-layer Model}

In two-dimensional turbulence, the energy flux $\Pi_E (k)$ and the enstrophy flux $\Pi_G (k)$ are constrained by
\begin{equation}
k^2 \Pi_E (k) - \Pi_G (k) < 0, \label{eq:ineq}
\end{equation}
for all wavenumbers outside of the forcing range. This  inequality was communicated to us by Danilov \cite{article:Tung:2005:1,submitted:Gkioulekas:3} and it implies that the contribution of the downscale energy cascade to the energy spectrum is overwhelmed by the contribution of the downscale enstrophy cascade and cannot be seen visually on a plot. This result was conjectured earlier by Smith \cite{article:Smith:2004} who debated the Tung-Orlando theory \cite{article:Orlando:2003} by arguing that the downscale energy cascade can never have enough flux to move the transition wavenumber $k_t$  into the inertial range. The obvious counterargument is that the two-layer model is a different dynamical system than the two-dimensional Navier-Stokes equations, and that  it is not obvious that the Danilov inequality cannot be violated in the two-layer model \cite{article:Tung:2005:1,article:Tung:2004}. After the debate with Smith \cite{article:Smith:2004,article:Tung:2004}, we identified  \cite{submitted:Gkioulekas:3} the essential mathematical difference between two-dimensional turbulence and the two-layer quasi-geostrophic model.

In the two-layer  model forcing is due to thermal heating, which injects energy directly into the baroclinic part of the total energy.  The two-layer fluid sits atop of an Ekman boundary layer near the ground, which introduces Ekman damping in the lower layer \cite{book:Holton:1979} but  \emph{not} in the upper layer. Following Salmon \cite{book:Salmon:1998}, one may then derive the governing equations for the model, which read:
\begin{align}
\pderiv{\gz_1}{t} &+ J(\gy_1, \gz_1+f) = -\frac{2f}{h} \gw + d_1\\
\pderiv{\gz_2}{t} &+ J(\gy_2, \gz_2+f) = +\frac{2f}{h} \gw + d_2 + 2e_2\\
\pderiv{T}{t} &+ \frac{1}{2}[J(\gy_1, T)+J(\gy_2, T)] = -\frac{N^2}{f}\gw + Q_0.
\end{align}
Here,  $\gz_1 = \del^2 \gy_1$ is the relative vorticity of the top layer and $\gz_2 = \del^2 \gy_2$  is the relative vorticity of the bottom layer, and $\gw$ is the vertical velocity. The temperature equation is situated between the two layers and it satisfies the geostrophic condition  $T=(2/h)(\gy_1-\gy_2)$ with $h$ the separation between the two layers. Furthermore, $f$ is the Coriolis term, $N$ is the \BVdude  frequency, and $Q_0$ is the thermal forcing on the temperature equation.  The dissipation terms include momentum dissipation of relative vorticity,  in each layer, and Ekman damping from the lower boundary layer, and they read:
\begin{align}
d_1 &= (-1)^{\gk+1}\nu \del^{2\gk} \gz_1\\
d_2 &= (-1)^{\gk+1}\nu \del^{2\gk} \gz_2 \\
e_2 &= -\nu_E \gz_2.
\end{align}
This model can be reduced to a coupled 2D-like system by employing the temperature equation to eliminate the vertical velocity $\gw$. This leads to the definition of the potential vorticity $q_1$ and $q_2$ as
\begin{align}
q_1 &=  \del^2 \gy_1 + f + \frac{k_R^2}{2}(\gy_2-\gy_1)\\
q_2 &=  \del^2 \gy_2 + f - \frac{k_R^2}{2}(\gy_2-\gy_1),
\end{align}
with $k_R \equiv 2\sqrt{2} f/(hN)$ the Rossby radius of deformation wavenumber. The governing equations for $q_1$ and  $q_2$ are shown to be
\begin{align}
\pderiv{q_1}{t} &+ J(\gy_1, q_1) = f_1 + d_1\\
\pderiv{q_2}{t} &+ J(\gy_2, q_2) = f_2 + d_2+ e_2,
\end{align}
Here $f_1$ and  $f_2$ is the thermal forcing on each layer given by $f_1 = -(k_R^2 Q)/(2f)$ and $f_2 =  (k_R^2 Q)/(2f)$ where $Q=(1/4)k_R^2 h Q_0$. 

The two inviscid quadratic invariants are the  energy $E$ and the total layer potential enstrophies $G_1$ and $G_2$ given by
\begin{align}
E &\equiv \snrm{\gy_1 q_1 + \gy_2 q_2} \\
G_1 &\equiv \snrm{q_1^2}, \quad G_2 \equiv\snrm{q_2^2}.
\end{align}
The energy and enstrophy spectra are defined as
\begin{align}
E(k) &\equiv \spec{\gy_1}{q_1}{k} + \spec{\gy_2}{q_2}{k},\\
G_1 (k) &\equiv \spec{q_1}{q_1}{k},\\
G_2 (k) &\equiv \spec{q_2}{q_2}{k},
\end{align}
and the total enstrophy spectrum $G (k)$ is $G(k) = G_1 (k) + G_2 (k)$.  It is also useful to distinguish between barotropic energy and baroclinic energy as follows: Let $\gy \equiv (\gy_1 + \gy_2)/2$ and $\gt \equiv (\gy_1 - \gy_2)/2$. So, $\gy_1 = \gy+\gt$ and $\gy_2 = \gy-\gt$. Now we define three spectra $E_K (k)$, $E_P (k)$, and $E_C (k)$ in terms of $\gy$ and $\gt$:
\begin{align}
E_K (k) &\equiv 2k^2 \spec{\gy}{\gy}{k}, \\
E_P (k) &\equiv 2(k^2 + k_R^2) \spec{\gt}{\gt}{k}, \\
E_C (k) &\equiv 2k^2 \spec{\gy}{\gt}{k}.
\end{align}
Here $E_K (k)$ is the barotropic energy spectrum and $E_P (k)$ the baroclinic energy spectrum. It is easy to show that the definitions are self-consistent, i.e. $E (k)=E_K (k)+E_P (k)$. The physical interpretation of $ E_C (k)$ is that it represents the difference in potential enstrophy distribution between the two layers, and it is given by
\begin{equation}
E_C (k) = \frac{G_1 (k) - G_2 (k)}{2(k^2+k_R^2)}.
\end{equation}

We have shown \cite{submitted:Gkioulekas:3} that it is the asymmetric presence of Ekman damping $e_2 = -\nu_E \gz_2$ on the bottom layer but not the top layer which causes the violation of the Danilov inequality \eqref{eq:ineq} in the two-layer model.  As a result, the top layer has more enstrophy than the bottom layer, as is realistic in the atmosphere, and provided that the difference in enstrophy between the two layers is large enough, the subleading downscale energy cascade will be observable in the energy spectrum.  If one artificially adds an identical  Ekman damping $e_1 = -\nu_E \gz_1$ in the upper layer  it can be easily shown that Danilov's inequality \eqref{eq:ineq} applies. In that case of symmetric dissipation, the subleading downscale energy cascade will be hidden by the dominant downscale enstrophy cascade.

For the case of asymmetric Ekman damping that we are considering here, we have shown \cite{submitted:Gkioulekas:3} that a \emph{sufficient} condition to \emph{satisfy} the Danilov inequality is
\begin{equation}
\nu_E < 4\nu k^{2p}_{\max} \fracp{k_{\max}}{ k_R}^2.
\end{equation}
Here $k_{\max}$ is either the truncation wavenumber in the numerical model, or, in the theoretical case of infinite resolutions, is the hyperviscosity dissipation wavenumber, beyond which the spectral enstrophy dissipation rate becomes negligible. Equivalently, a \emph{necessary} condition to \emph{violate} Danilov's inequality is 
 \begin{equation}
 \nu_E > 4\nu k^{2p}_{\max}\fracp{k_{\max}}{k_R}^2.
 \end{equation}

We have also derived \cite{submitted:Gkioulekas:3} a \emph{necessary and sufficient condition} for violating the Danilov inequality.  However, the price that must be paid for doing so is that the condition is \emph{uncontrolled}. By this, we mean that the condition has the form 
\begin{equation}
\nu_E k_R^2 > \gL \nu k_{\max}^{2p+2},
\end{equation}
 but it is not possible to find a universal value for $\gL$ that will \emph{always} work.  We have shown that the necessary requirement needed to \emph{have} a sufficient condition for violating the Danilov inequality at wavenumber $k$ is
\begin{equation}
G_1 (q)- (1+4(q/k_R)^2) G_2 (q) > 2q^2 E_K (q),
\end{equation}
for all $q$ such that $k<q<k_{max}$.

It should be noted that the simulation of Tung and Orlando \cite{article:Orlando:2003} has already shown that it is possible to have an observable downscale energy cascade. The only issue that required clarification was to understand why it happens in the two-layer model but not in two-dimensional turbulence, when the  value of Ekman damping is larger than the subgrid hyperdiffusion.

\subsection{Surface Quasi-geostrophic Models}

Although the troposphere is thin, it is anisotropic in the vertical.  In meteorological jargon it is said that the dynamics of the troposphere is baroclinic.  Baroclinic instability in the troposphere has traditionally been studied using two-layer models \cite{book:Holton:1979}.  We have discussed in previous sections that a prerequisite of a model of the large-scale turbulence in the troposphere is the ability to model in some way its baroclinicity.  In the case of two-layer model, it is essential that the two layers of the model do not behave in a similar way and that there is a large vertical difference.  Otherwise, the model will degenerate into two two-dimensional  turbulence layers, and we know that for each layer Danilov's inequality would hold and no transition in the Nastrom-Gage spectrum can be obtained.

The troposphere's vertical anisotropy is actually more extreme than that in a two-layer model.  Because of fast mixing in the vertical, the potential vorticity is homogenized over the depth of the troposphere, thus creating a surface of discontinuity at the tropopause, which separates a zero-potential vorticity troposphere from a high potential vorticity stratosphere.  The instability arises from the interaction of the tropopause and the ground has been studied using the so-called Eady model \cite{article:Eady:1949}, which is actually simpler than the two-layer model.  The nonlinear version of the Eady model was studied by Blumen \cite{article:Blumen:1978,article:Blumen:1978:1} and by Held and Pierrehumpert \cite{article:Swanson:1994,article:Swanson:1995}.  Although we have not yet completed a simulation of the Nastrom-Gage spectrum using such a model, called 2-surface sQG, by Muraki and Hakim \cite{article:Hakim:2001}, we have previously derived some theoretical results \cite{article:Orlando:2003:1,submitted:Gkioulekas:3} where we have  shown that for large wavenumbers $k$ the energy spectrum should behave like $k^{-5/3}$.   Such a downscale energy flux remains finite even as viscosity is reduced to zero from above \cite{article:Orlando:2003:1}.  It then appears that in such a model the mesoscale spectrum of $-5/3$ can be simulated, at least the upper part (the longer scales) of the $-5/3$ spectrum.  As pointed out previously \cite{submitted:Gkioulekas:3}, for the shorter wave part of the mesoscale spectrum, especially for length scales shorter than 100km, QG scaling fails and other motion (such as stratified turbulence, and three-dimensional  turbulence) takes over.  Nevertheless, dimensional analysis would still yield a $-5/3$ energy spectrum as long as there is a downscale energy flux through this range.

\section{The Method of Spectral Reduction}

In standard numerical simulations of two-dimensional turbulence and quasi-geostrophic turbulence we have to trade-off numerical resolution with run-time.  Many of the really interesting questions require both very high numerical resolution and very long run-time simultaneously.  An interesting option for outflanking this problem is the method of Spectral Reduction \cite{article:Bowman:1996:1,article:Morrison:1999}. This method reduces the governing equations to a small number of wavenumber shells with each shell further subdivided into a small number of sectors.  In this sense, one can say that spectral reduction is a method for generating a ``realistic'' shell model for the nonlinear dynamics of the Navier-Stokes equations. 

 It has been known for some time that shell models can reproduce the intermittency corrections of three-dimensional turbulence in agreement with experimental measurements \cite{book:Vulpiani:1998,article:Biferale:2003}. It still remains a mystery why they work so well, especially given that there is no direct mathematical relationship between the models themselves and the Navier-Stokes equations. However, Spectral Reduction differs from shell models in one crucially important way: It is an approximation  which is applied directly on the Navier-Stokes equations and thus the resulting ODE model is a realistic approximation of the shell interactions rather than an ad hoc cascade model.  Furthermore, it preserves the conservation laws of the original equations.  Apparently, the   method exploits some redundancy that is introduced into the spectral representation of the velocity field after taking a time average. This allows the discretization to converge extremely rapidly, \emph{as long as the results are  time-averaged}.

Because the method is not well known, we give a brief technical description. Let $a (\bfx,t)$ and $b (\bfx,t)$ be two fields with Fourier transforms $\hat a_\bfk$ and $\hat b_\bfk$, and let $\cV_{\bfk\bfp\bfq}$ be the kernel of the Jacobian operator defined as 
\begin{equation}
\hat J_\bfk = \int d\bfp \int d\bfq \; \cV_{\bfk\bfp\bfq} \hat a_\bfp \hat b^{\ast}_\bfq,
\end{equation}
where $\hat J_\bfk$ is the Fourier transform of $J(a,b)$. 
Let $\cD \subseteq \bbR^2 - \{\bfzero\}$ be a bounded  continuous domain of wavenumbers,  excluding a neighborhood of $\bfzero$, that has been partitioned into $N$ connected regions $\cR_n \subset \cD$. Each region has area $A_n$ and a center-of-mass wavenumber $K_n$ defined as
\begin{align}
A_n &= \int_{\cR_n} d\bfk\\
K_n &= \frac{1}{A_n} \int_{\cR_n} \nrm{\bfk} \; d\bfk.
\end{align}
Next, we associate with each region an average vorticity $Z_n$ given by
\begin{equation}
Z_n = \frac{1}{A_n}\int_{\cR_n} \hat \gz_\bfk \; d\bfk,
\end{equation}
and using  the ad-hoc ``approximation'' $\hat \gz_\bfk = Z_n, \forall \bfk\in \cR_n$ we obtain the following governing equation for $Z_n$:
\begin{equation}
\pderiv{Z_n}{t} + J_n = D_n Z_n + F_n, 
\end{equation}
where $D_n$ represents the dissipation operators and $F_n$ the forcing term, and they are given by
\begin{align}
D_n &= \frac{1}{A_n}\int_{\cR_n} (-\nu \nrm{\bfk}^{2\gk} - \gb \nrm{\bfk}^{-2m}) \; d\bfk\\
F_n &=\frac{1}{A_n}\int_{\cR_n} \hat f_\bfk \; d\bfk.
\end{align}
The non-linearity $J(\psi,\gz)$ is in the term $J_n$, which is further approximated by
\begin{align}
J_n &= \sum_{a = 1}^N \sum_{b = 1}^N V_{nab}\frac{Z_a Z^{\ast}_b}{K_b^2}\\
V_{nab} &= \frac{1}{A_n A_a A_b} \left[\int_{\cR_n} d\bfk \int_{\cR_a} d\bfp \int_{\cR_b} d\bfq \cV_{\bfk\bfp\bfq} \right].
\end{align}
The conserved energy $E$ and enstrophy $G$ for this system are given by
\begin{align}
E &= \sum_{n= 1}^N \frac{|Z_n|^2}{K_n^2} A_n  \\
G &=  \sum_{n= 1}^N  | Z_n|^2  A_n.
\end{align}

It has been demonstrated that spectral reduction can reproduce the enstrophy cascade of two-dimensional turbulence \cite{article:Morrison:1999}, and furthermore in agreement  with the predictions of  Falkovich and Lebedev \cite{article:Lebedev:1994,article:Lebedev:1994:1}. This is very strong evidence that the method is effective in characterizing the statistical features of the enstrophy cascade. Recall that the possibility of intermittency corrections for the downscale enstrophy cascade has been rigorously ruled out \cite{article:Eyink:2001}, so these numerical predictions are in agreement with a rigorous mathematical result. The shortcoming of spectral reduction is that it does not calculate correctly the absolute equilibrium energy spectrum of two-dimensional turbulence \cite{article:Morrison:2001}.  It has been shown \cite{article:Morrison:2001} that this can be corrected if one uses instead the rescaled equations 
\begin{equation}
\frac{A_0}{A_n}\pderiv{Z_n}{t} + J_n = D_n Z_n + F_n.
\end{equation}
However, the rescaling comes with the price that for numerical stability it is now necessary to use a smaller timestep.  It has also been found that it is necessary to use the rescaled equations to reproduce the inverse energy cascade \cite{article:Morrison:2001}.  It is not well-understood why  the enstrophy cascade can be obtained without rescaling and, more importantly, why the inverse energy cascade cannot.  We conjecture that the reason is that the inverse energy cascade is very close to absolute equilibrium \cite{article:Procaccia:2002} whereas the enstrophy cascade is not.  In any event, even with the rescaled spectral reduction, we still benefit in computation time from not having to work with very large data sets and Fourier transforms at every step.

  From a mathematical perspective, the main idea behind Spectral Reduction is having a method for approximating nonlinear terms that can be expressed in terms of  Jacobian operators $J(a,b)$.  This means that it is possible in principle to extend the method to any physical model where the nonlinear terms can be written terms of Jacobians. This is important because it is known that 3D  large-scale flows conserve potential vorticity layerwise \cite{article:Charney:1947:1}.  For each layer of the numerical model, the nonlinear term is in  the form of a Jacobian in two dimensions. This covers the two-layer model but it also allows the possibility to consider models with many layers.  Such models would be very difficult to examine with direct numerical simulation.

\section{Conclusions}

We would now like to summarize the main points of this paper.  We have seen that two-dimensional turbulence has been investigated numerically to considerable detail.  Both the downscale enstrophy cascade and the inverse energy cascade have been successfully observed.  However, neither cascade is as robust as the downscale energy cascade of three-dimensional turbulence.  The disruption of cascades in two-dimensional turbulence is associated physically with the emergence of long-lived coherent structures.  It is not really understood why these coherent structures emerge in two-dimensional turbulence but not in three-dimensional turbulence.  However, we have reviewed   some significant breakthroughs in understanding the particulars of the cascades of two-dimensional turbulence, when these cascades are not disrupted.

The situation becomes even more interesting for the case of flows that are approximately two-dimensional, and especially in the context of understanding the Nastrom Gage energy spectrum of the atmosphere.  We have reviewed some of the proposed theories, and discussed more extensively the Tung-Orlando  theory. All that can be said with certainty is that the work to date in this direction raises more questions than it answers!  Consequently an open mind is needed to make further progress.

Although there is considerable interest in two-dimensional turbulence on the one hand, and in General Circulation Models on the other hand,  there is relatively limited interest in the  theoretical understanding  of simpler models in between these two extremes, such as, for example, the two-layer quasi-geostrophic model.  Finite layer quasi-geostrophic models have the advantage that they are possibly within range of theoretical analysis using tools that have proved themselves in studies of two-dimensional turbulence.  It is worthwhile to study these models for two reasons:  first, ``because they are there'';  second, because, as  Held \cite{article:Held:2005} has pointed out, genuine understanding arises from a comprehensive study of the entire spectrum of models, from the simplest to the most realistic. 

We have concluded our review  with a brief discussion of the method of spectral reduction.  We suggest that this method could prove effective in investigating the phenomenology of quasi-geostrophic models.

\section*{Acknowledgement}

It is a pleasure to thank  Sergey Danilov, Gregory Eyink, and Victor L'vov for  e-mail correspondance and discussions and  Sergey Nazarenko, Dwight Barkley, Robert Kerr, Oleg Zaboronski, Peter McClintock, Grisha Volovik, Martin Wilkens   for organizing the excellent December meeting of the Warwick Turbulence Symposium 2005-2006. The research is supported in part by the National Science Foundation, under grants DMS-03-27658 and ATM-01-32727.

\bibliography{references,references-submit}
\bibliographystyle{spmpsci}
\end{document}